\documentclass[%
 jap,
 jmp,%
 amsmath,amssymb,
 reprint,%
]{revtex4-1}

\usepackage{graphicx}
\usepackage{dcolumn}
\usepackage{bm}
\usepackage{amsmath}
\usepackage{ amssymb }

\begin{document}





\title{Magneto-Plasmonic Nanoantennas: Basics and Applications (Review)}
\author{Ivan S. Maksymov}
\email{ivan.maksymov@rmit.edu.au} 
\affiliation{ARC Centre of Excellence for Nanoscale BioPhotonics, School of Applied Sciences, RMIT University, Melbourne, VIC 3001, Australia}

\date{\today}

\begin{abstract}
Plasmonic nanoantennas is a hot and rapidly expanding research field. Here we overview basic operating principles and applications of novel magneto-plasmonic nanoantennas, which are made of ferromagnetic metals and driven not only by light, but also by external magnetic fields. We demonstrate that magneto-plasmonic nanoantennas enhance the magneto-optical effects, which introduces additional degrees of freedom in the control of light at the nano-scale. This property is used in conceptually new devices such as magneto-plasmonic rulers, ultra-sensitive biosensors, one-way subwavelength waveguides and extraordinary optical transmission structures, as well as in novel biomedical imaging modalities. We also point out that in certain cases 'non-optical' ferromagnetic nanostructures may operate as magneto-plasmonic nanoantennas. This undesigned extra functionality capitalises on established optical characterisation techniques of magnetic nanomaterials and it may be useful for the integration of nanophotonics and nanomagnetism on a single chip. 
   
\end{abstract}


\maketitle 

\section{Introduction}

\begin{figure}[t]
\centering\includegraphics[width=8.0cm]{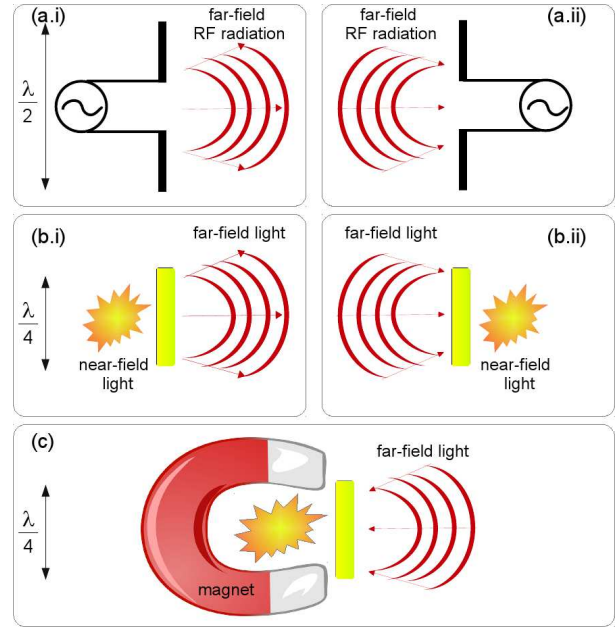}
\caption{ (a) Illustration of emission (left) and reception (right) of radio waves by a dipole RF antenna. (b) Illustration of emission (left) and reception (right) of light by a dipole plasmonic nanoantenna. In all Panels, $\lambda$ denotes the wavelength of the incident radio waves or light in free space. The double-headed arrows indicate the dimensions of the antennas in comparison with $\lambda$. (c) Optical properties of magneto-plasmonic nanoantennas are similar to those of non-magnetic nanoantenas. However, the use of magnetic constituent materials and external magnetic fields adds new degrees of freedom in the control of light at the nano-scale, which allows developing novel devices with unique properties (see the main text).}
\label{fig1}
\end{figure}

Plasmonic nanoantennas emit, receive and, more generally, control light with nano-scale (sub wavelength) elements, whose size is much smaller than the wavelength of incident light \cite{Nov11, Kra13}. They are made of single metal nanoparticles (usually gold or silver) or their constellations, and their design visually resembles the existing structures of RF antennas \cite{Balanis} such as dipole antennas [Fig.~\ref{fig1}(a)].

However, the operating principles of plasmonic nanoantennas and RF antennas are different. The response of nanoantennas to incident light is dictated by collective electron oscillations -- plasmons \cite{Bar03}. Plasmons make it possible to control light with subwavelength structures, which is not readily possible with RF antennas whose dimensions are comparable with the wavelength of radio waves. Moreover, nanoantennas not only control light similar to radio waves [Fig.~\ref{fig1}(b)], but they also locally enhance optical intensity by many orders of magnitude \cite{Nov11, Kra13}. This effect is achievable because of a strong local field confinement near the metal surface of the nanoantenna, and it is used to enhance the extremely small nonlinear optical response of nanoscale materials up to the level achievable with macroscopic nonlinear crystals and optical fibres \cite{Kra13}.

Different aspects of plasmonic nanoantennas were discussed in detail in \cite{Bra09, Alu09, Nov11, Gia11, Bon12, Ber12, Agi12, Bia12, Mak12, Che12, Kra13, Alu13, Agi13, Cha13, Mon14, Bor14}. However, the research direction of nanoantennas is so broad and rapidly expanding that it is virtually impossible to write a comprehensive review covering all aspects. Thus, in this review paper we survey the operating principles and applications of novel \textit{magneto-plasmonic nanoantennas}.

In contrast to conventional (non-magnetic) plasmonic nanoantennas, the constituent materials of magneto-plasmonic nanoantennas are ferromagnetic metals such as nickel, cobalt, iron or their alloys. Alternatively, magneto-plasmonic nanoantennas can be made of a non-magnetic metal combined with a magnetic material, which can either be conducting or insulating. In addition, external magnetic fields need to be applied to magneto-plasmonic nanoantennas to saturate the magnetisation along one of the coordinate directions [Fig.~\ref{fig1}(c)]. It is noteworthy that very weak magneto-optical activity can also be detected in non-magnetic plasmonic nanoantennas \cite{Sep10, Pin13, Tub13}. However, such devices will not be considered in this work because their operation requires impractically large magnetic fields. 

The time is definitely ripe to overview the breakthroughs in the field of magneto-plasmonic nanoantennas. Although different aspects of nano-scale magneto-plasmonics have been discussed in detail in \cite{Tem12, Arm13, Ino13, Arm14, Mak15, Luk16}, magneto-plasmonic nanoantennas and their applications were not in the focus of the previous works. Moreover, there has been a big progress in the last two years leading to significant results that are scattered in the literature. Consequently, in this review paper we will catch up to the advances made in this research field by presenting the reader with information on basic operating principles of magneto-plasmonic nanoantennas and their applications in photonics and biomedical imaging.

\section{Magneto-plasmonic nanoparticles} \label{Magneto-plasmonic nanoparticles}

Although research on magneto-plasmonic nanoantennas is rapidly maturing, historically magneto-plasmonic nanoparticles made of ferromagnetic metals or insulating magnetic materials combined with noble metals were investigated first. Moreover, a small gold nanoparticle was the key element of the first optical nanoantenna \cite{Nov}. Thus, we start our discussion with an overview of magneto-plasmonic nanoparticles. 

The problem of light scattering by a spherical nanoparticle of arbitrary diameter and dielectric permittivity is exactly soluble by using the Mie theory \cite{19,20}. Whereas the Mie theory can be extended to calculate the scattering by homogeneous magnetic spheres described by their magnetic permeability \cite{20}, the problem of light scattering by a sphere magnetised along a certain coordinate direction is more difficult. This is because the dielectric permittivity becomes a tensor describing the interaction between light and external magnetic fields (or the internal magnetisation of the medium) \cite{2}.

By considering the off-diagonal permittivity tensor elements in the solution of the Maxwell's equations for the electromagnetic fields, one can calculate the magneto-optical Kerr and Faraday effects contribution to the optical response of a spherical nanoparticle \cite{22, 23}. The magneto-optical Kerr effect in a homogeneous cobalt sphere with the radius of $260$~nm was calculated using a perturbation technique in \cite{23}. This radius is larger than the critical radius for cobalt ($a_{\rm{cr}} = 30$~nm \cite{Kak05}), which implies that the chosen sphere is no more a monomagnetic domain unless a static magnetic field is applied to enforce a single-domain structure and control the direction of the magnetisation inside the sphere. It was demonstrated that the intensity of the forward- and back-scattered light is unaffected by the effect of magnetisation because in the perturbation approach the corrections to the electromagnetic fields due to the magnetisation are very small. However, the intensity is essentially modified for the scattering angle $\theta = \frac{\pi}{4}$, which offers the opportunity to dynamically tune the light scattering pattern by changing the magnetisation direction.

Next, we discuss the magneto-optical response of nanoparticles consisting of a magnetic material and a noble metal \cite{Kos98, Abe05, Li05, Smi05, Jai09, Bar09, Wan11, Toa14, Bar15, Luk16}. Plasmon-enhanced magneto-optical Faraday rotation was demonstrated in a colloidal solution of physically conjoined nanoparticle pairs composed of a spinel ferrite CoFe$_2$O$_4$ and silver \cite{Li05}. Spinel ferrites are a class of compounds of general formula MFe$_2$O$_4$ with M = Mn, Co, Ni, Zn, Mg, etc., which are of great interest for their remarkable magnetic and optical properties \cite{Car09, Har09}. It was shown that at certain wavelengths of incident light the magneto-plasmonic nanoparticle pairs exhibit a significantly enhanced magneto-optical response as compared with single CoFe$_2$O$_4$ nanoparticles that do not support plasmon modes. However, the experimental results presented in \cite{Li05} lack of full magneto-optical spectra, which makes it difficult to fully understand the nature of the plasmon-enhanced magneto-optical effect. 

\begin{figure}[t]
\centering\includegraphics[width=8.0cm]{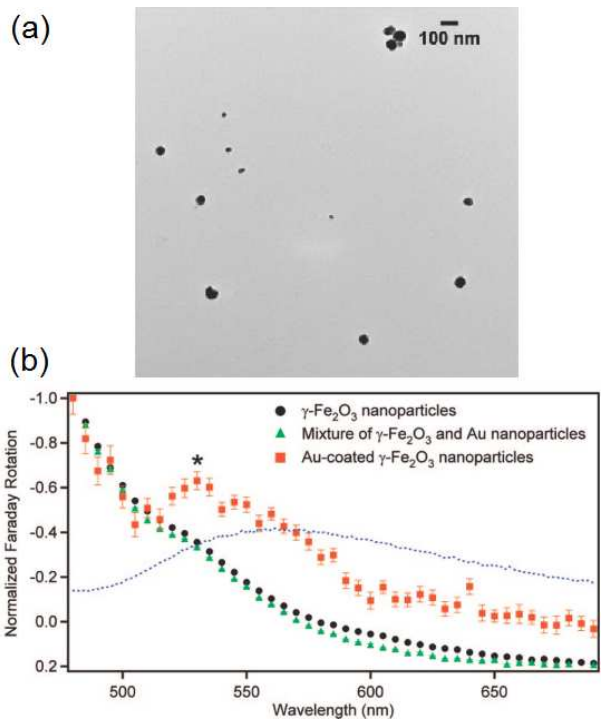}
\caption{ (a) Transmission electron microscopy image of gold-coated $\gamma$-Fe$_2$O$_3$ nanoparticles. (b) Normalised Faraday rotation spectra of $\gamma$-Fe$_2$O$_3$ nanoparticles, gold-coated $\gamma$-Fe$_2$O$_3$ nanoparticles, and a mixture of $\gamma$-Fe$_2$O$_3$ and gold nanoparticles. The absorption spectrum showing the plasmon resonance band of the gold-coated $\gamma$-Fe$_2$O$_3$ nanoparticles is indicated by the dotted blue curve. The * sign represents the position of the absorption band edge in $\gamma$-Fe$_2$O$_3$. Adapted from \cite{Jai09}.}  
\label{fig3}
\end{figure}

Full magneto-optical spectra and a comprehensive discussion of the origin of the plasmon-enhanced magneto-optical effect were presented in \cite{Jai09}, in which enhanced optical Faraday rotation was reported in gold-coated maghemite ($\gamma$-Fe$_2$O$_3$) nanoparticles [Fig.~\ref{fig3}(a)]. The Faraday rotation spectrum measured in the $480-690$~nm spectral range shows a peak at about $530$ nm, which is not present in either uncoated maghemite nanoparticles or solid gold nanoparticles [Fig.~\ref{fig3}(b)]. In fact this peak corresponds to an intrinsic electronic transition in the maghemite nanoparticles and is consistent with a near-field enhancement of the Faraday rotation resulting from the spectral overlap of the surface plasmon resonance in the gold with the electronic transition in maghemite.

Light absorption losses in single metal nanoparticles or metal-coated nanoparticles are unavoidable because all metals absorb light. However, absorption losses can be significantly reduced by using active core-shell nanoparticles combining metals with a gain material \cite{Bar15}. Such nanoparticles are also called 'spasers'. Thanks to the gain, a guided mode in an array of spasers may exhibit high values of the Faraday rotation and propagation length in the array.

The optical response of magneto-plasmonic nanoparticles can be controlled by engineering their shape and size. Although the shape and size can be controlled at will by using colloidal techniques \cite{Jai09}, modern techniques for fabrication of ferromagnetic nanostructures \cite{Ade08} may be more suitable for the geometrical tuning. For example, magneto-plasmonic nanoring nanoparticles were introduced in \cite{Fen14} as novel metamaterials with tunable magneto-optical activity in a wide spectral range. Such metamaterials may be useful in optical data storage, miniaturised magneto-optic devices, as well as in optical sensing and imaging of magnetic fields and magnetic domain structures \cite{Bar09, Mak15}.

Finally, nanoparticles combining tunable optical response with a strong magnetic response are often employed in therapy and diagnostics (theranostics) \cite{Lar07, Bar11}. In those research direction, magnetic nanoparticles are especially attractive because external magnetic fields can be used to remotely guide the delivery of nanoparticles inside the human body. We will continue this discussion in Section \ref{imaging}.

\section{Magneto-plasmonic nanoantennas} \label{Magneto-plasmonic nanoantennas}

\subsection{Origin of the concept}

\begin{figure*}[t]
\centering\includegraphics[width=12cm]{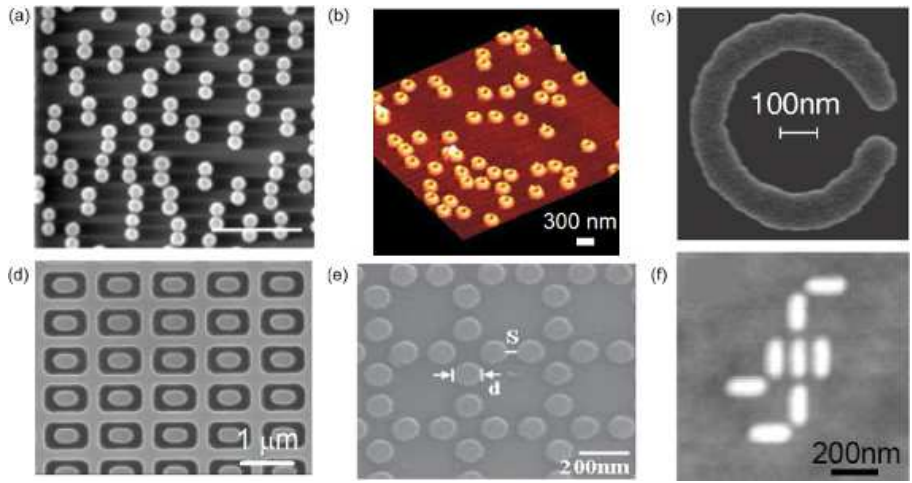}
\caption{ Images of representative magnetic nanostructures that support plasmonic excitations and operate as magneto-plasmonic nanoantennas when they are illuminated by visible or near-infrared light. (a) Nickel discs pairs with the gap size of $10$~nm used as a magneto-plasmonic ruler \cite{Zub15}. The scale bar is $1$~$\mu$m. (b) Atomic Force Microscopy profile image of gold-cobalt-gold nano-rings used as broadband and tunable magneto-plasmonic nanoantennas \cite{Fen14}. (c) Permalloy (Ni$_{80}$Fe$_{20}$) gapped nano-rings used in magnetic data storage \cite{Li12}. (d, e) Permalloy anti-rings \cite{Din13} and dots arranged into the octagonal lattice used as magnonic crystals \cite{Sah13}. (f) Iron nanostructure of a magnetic quantum cellular automata \cite{Gro10}. Note that only the structures in Panels (a) and (b) were designed to operate as optical devices. Although the structures in the remaining Panels were not designed as such, their metallic nanostructures allow employing them as optical devices, which opens up opportunities for the integration of nanophotonics and nano-magnetism \cite{NKos13, Mak15}.}  
\label{fig4}
\end{figure*}

In contrast to magneto-plasmonic nanoparticles fabricated by using colloidal techniques \cite{Jai09}, ferromagnetic metal nanostructures operating as magneto-plasmonic nanoantennas \cite{Zub15, Fen14, Li12, Din13, Sah13, Gro10} (Fig.~\ref{fig4}) are engineered on top of dielectric substrates by using fabrication techniques that may be compatible with complementary metal-oxide-semiconductor (CMOS) manufacturing process (see, e.g., \cite{Ade08}). However, not all structures in Fig.~\ref{fig4} were designed to operate as optical devices, which implies that ferromagnetic metal nanostructures are not exclusively a product of advances in nanophotonics. 

Indeed, the origin of many ferromagnetic nanostructures can be traced back to the advances in 'non-optical' areas of magnonics \cite{Kru10, Mak15_Kostylev} and spintronics \cite{Hir14}, in which the representative devices are magnetic multilayers \cite{Mak15_Kostylev}, magnonic crystals \cite{Kru10}, spin torque nano-oscillators \cite{Kim12}, and magnetic quantum cellular automata \cite{Cow00}. As can be seen from Fig.~\ref{fig4}(c-f), the design of such devices is ideally suitable for the operation in the optical regime because the metal features of their nanostructures support plasmon modes \cite{Arm13, NKos13, Mak15}.  

Whereas ferromagnetic metal discs, ellipses, and rings are easily identified as plasmonic nanoantennas by readers familiar with nanoplasmonics, in magnonics and spintronics the concept of magneto-plasmonic nanoantenna remains relatively unknown \cite{Ele07, NKos13, Uch15, Mor15}. However, very often the magnetic properties of magnonic and spintronic devices are probed with light, which is the case of the Brillouin Light Scattering (BLS) spectroscopy and magneto-optical Kerr effect (MOKE) magnetometry \cite{San96, Vav00, Gub05, Seb15}. It was suggested that the nanoantenna-like behaviour of ferromagnetic metal magnonic crystals and similar structures may affects experimental results obtained with these techniques, or, if properly understood, it may be used to improve the resolution of these techniques \cite{NKos13, Mak15}. Most significantly, the nanoantenna-like behaviour may bridge the gap between nanoplasmonics and magnonics (or spintronics), and lead to the development of new hybrid magneto-photonic nanotechnologies \cite{Mak15}.

The existing gap between nanoplasmonics and magnonics cannot be bridged without the use of materials common to both technologies. The application of plasmonic ferromagnetic metals, such as nickel, iron and cobalt, is limited in magnonics and spintronics \cite{Mak15_Kostylev}. Instead of them magnonics and spintronics often rely on Permalloy (Ni$_{80}$Fe$_{20}$) because this magnetic alloy has the optimum combination of microwave magnetic properties such as the vanishing magnetic anisotropy and the smallest magnetic (Gilbert) damping among ferromagnetic metals \cite{Mak15_Kostylev}. Remarkably, in recent works \cite{NKos13, Ber15, Mac15_ACS} it was shown that Permalloy also exhibits feasible plasmonic properties. 

Finally, it is noteworthy that Permalloy and other ferromagnetic metal nanostructures may have technologically important non-magnetic gold, silver, platinum, palladium or tantalum layers \cite{Ade08, Mak15_Kostylev}. From the optical point of view, these layers may be useful. However, one should bear in mind that these layers also adversely affect microwave magnetic properties of ferromagnetic nanostructures \cite{Mak15_Kostylev}.

\subsection{Design and operating principles}

From the purely optical point of view, the design of magneto-plasmonic nanoantennas may rely on the same principles used in non-magnetic nanoantennas. Here we are interested in simple nanoantenna architectures such as discs, ellipses and nanorods. Although more complex designs of non-magnetic nanoantennas are possible (e.g., the Yagi-Uda nanoantennas \cite{Mak12}), absorption losses in ferromagnetic metals are larger than in gold and silver, and thus they make it challenging to create efficient multi-element magneto-plasmonic nanoantennas.

The length of the dipole RF antenna is approximately half the wavelength of the incident radio waves [Fig.~\ref{fig1}(a)]. However, the length of the dipole plasmonic nanoantenna made of gold or silver is smaller than the wavelength $\lambda$ of incident light in free space [Fig.~\ref{fig1}(b)]. This is because the penetration depth of light into metals and the excitation of plasmons limit the possibility to directly downscale the RF antenna constructions \cite{Nov11, Kra13}.  As a rule of thumb, the length of the dipole nanoantenna equals half of an effective wavelength $\lambda_{\rm{eff}} = a + b\frac{\lambda}{\lambda_{\rm{p}}}$, being $a$ and $b$ some coefficients with dimensions of lengths and $\lambda_{\rm{p}}$ the plasma wavelength for the constituent metal \cite{Nov11}. This rule works very well in the case of other plasmonic metals, including ferromagnetic metals.

In addition to the effective wavelength rule, one has to take into account the fact that the plasmon response of some ferromagnetic metals such as nickel and iron is strongly affected by the presence of broad interband transition backgrounds in the spectral range of interest for nanoplasmonics \cite{Mel03, Mel12, Kat15}. In conventional non-magnetic nanoplasmonics, these features are regarded as parasitic because they increase light absorption losses \cite{Pak11}. Nickel has spectrally localised interband transitions at around $265$~nm \cite{Pir14}. The interaction between plasmons and interband transitions in nickel can be taken into account in the design as two coupled harmonic oscillators. The response of the plasmon-interband coupled system exhibits energy anti-crossing that is typically observed in strongly coupled systems.

Most significantly, apart from the purely optical considerations, the design of magneto-plasmonic nanoantennas has to take into account their magnetic response, which can be done by solving static and dynamic micromagnetics problems, such as solutions for the exchange energy, anisotropies, demagnetising fields, and the Landau-Lifshitz-Gilbert equation of magnetisation motion \cite{Sch02, Vav04, Nas11, Mac13, Mac13_1, Lod14, Mak15_Kostylev}. Because analytic approaches are only possible in some particular cases (e.g., ellipsoid structures), sophisticated numerical approaches need to be used. Thus, in the design of magneto-plasmonic nanoantennas one should combine theoretical and experimental knowledge of nanoplasmonics and nanomagnetism.

\begin{figure}[t]
\centering\includegraphics[width=6.5cm]{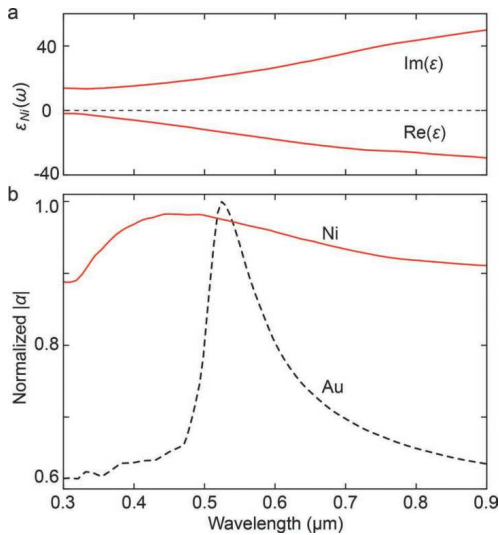}
\caption{(a) Dielectric permittivity of nickel as a function of the wavelength of light. (b) Comparison of the theoretical polarisability of nickel and gold spheres. Adapted from \cite{Che11}.} 
\label{fig5}
\end{figure}

For example, this combined approach was employed to design magneto-plasmonic nanoantennas consisting of nickel discs \cite{Che11, Bon11}, which support dipolar plasmon modes. Figure~\ref{fig5}(a) shows the dependence of the dielectric permittivity of nickel on the wavelength of incident light. Figure~\ref{fig5}(b) shows the theoretical polarisability of nickel and gold spheres. (Theoretical analysis of small spheres is relatively simple and it allows projecting the result onto more complex nanoantenna designs.) Compared to the polarisability of the gold sphere of the same diameter, the resonance of the nickel sphere is weaker due to larger absorption losses. However, one can clearly see the plasmon resonance in the spectrum. The magnetic response of the nickel nanoantenna was also confirmed by measuring the magneto-optical Kerr effect (MOKE) as a function of the external static field \cite{Arm13}. Those results (Fig. $2$ in \cite{Che11}) revealed hysteresis loops typical of ferromagnetic nanostuctures \cite{Cow99}.

\begin{figure}[t]
\centering\includegraphics[width=8.0cm]{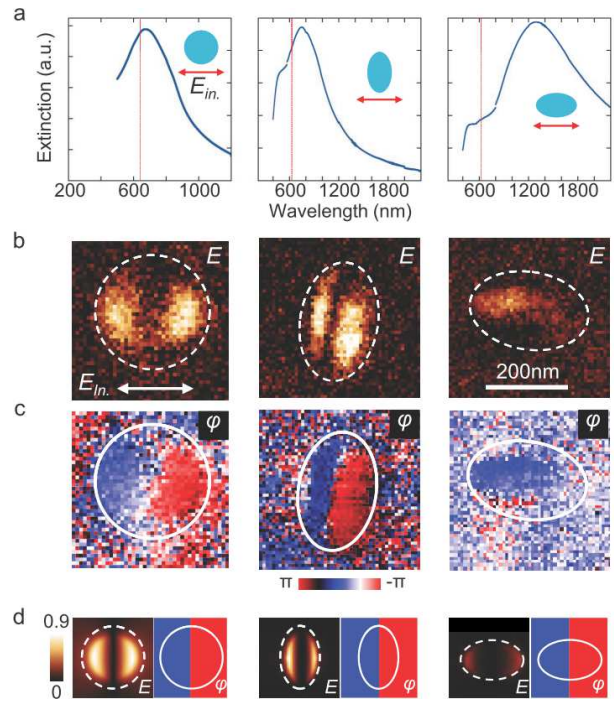}
\caption{ (a) Normalised extinction spectra of $200$~nm diameter nickel discs (left), nickel ellipses with polarisation along the short axis ($190$~nm, center), and nickel ellipses with polarisation along the long axis ($300$~nm, right). The red lines in the spectra mark the near-field imaging wavelength. (b) Near-field amplitude images. The arrow and bar denote the polarisation of incident laser beam and the scale of the images, respectively. (c) Near-field phase images. (d) Calculated near-field amplitude and phase maps at $633$ nm. The white circles in the near-field amplitude and phase images indicate the nickel nanoantenna. Adapted from \cite{Che11}.} 
\label{fig6}
\end{figure}

The far-field extinction spectrum [Fig.~\ref{fig6}(a)] of the $200$~nm discs exhibits a resonance peak at around $\lambda=650$~nm. Near-field imaging with a scattering-type scanning near-field optical microscope (\textit{s}-SNOM) operating at $\lambda=633$~nm was used to image the amplitude and phase of the out-of-plane near-field component of the plasmon modes [Fig.~\ref{fig6}(b, c)]. In the amplitude image, one can see two bright spots aligned along the polarisation direction. Similar to non-magnetic gold nano-discs \cite{Est08}, oscillations of the amplitude of these spots have the phase difference of $180^{\rm{o}}$. Numerical simulations of the out-of-plane components of the near-field of the nickel nano-discs confirmed the dipolar origin of the plasmon modes observed in the experiment [Fig.~\ref{fig6}(d)].

For the elliptical discs (the middle and the rightmost columns in Fig.~\ref{fig6}) both experiment and simulations revealed a plasmon response that strongly depends on the orientation of the ellipse. For a polarisation parallel to the short axis of the ellipse ($190$~nm), the far-field measurement revealed a resonance at around $\lambda=700$~nm. However, for a polarisation along the long axis of the ellipse ($300$~nm), the peak in the far-field spectrum was observed at $1500$~nm. As in the case of nano-discs, the near-field amplitude and phase images for the ellipses were taken at $\lambda=633$~nm. These results revealed a dipolar mode for the polarisation along the short axis, for which the imaging wavelength is close to the plasmon resonance. This result was also confirmed by simulations. Although in measurements of the elliptical nanoantennas rotated for $90^{\rm{o}}$ the imaging wavelength $\lambda=633$~nm was off-resonance and thus the mode picture could not be taken, simulations confirmed that the longitudinal plasmon fields are much weaker than the transverse fields on resonance.

We also highlight the result presented in \cite{Ber15} that reports an unusual enhancement of the magneto-optical effects in Permalloy disc nanoantennas with the diameter $D<400$~nm. The separation between the discs was chosen such that theoretical modification of the magneto-optical effects due to the optical coupling between the discs was negligibly small. Magnetic hysteresis loops of the discs were measured as relative diffracted light intensity change $\Delta I/I$ as a function of the external static magnetic field. These curves revealed ordinary features including the pinched shape of the loops corresponding to low field vortex states \cite{Cow99}.

\begin{figure}[t]
\centering\includegraphics[width=7.0cm]{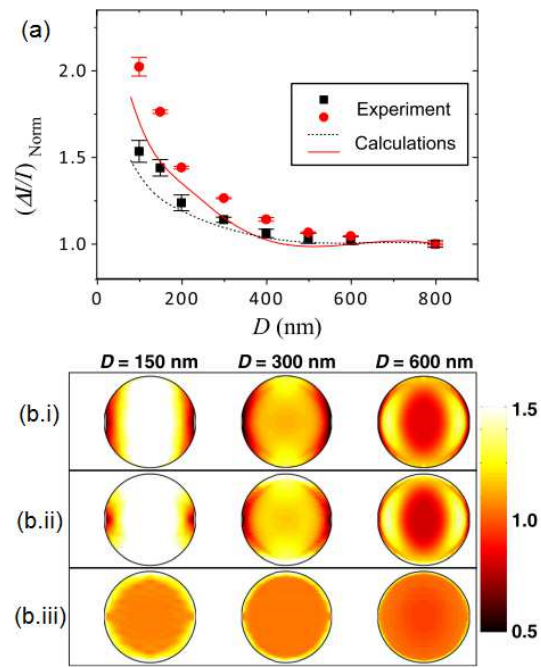}
\caption{Normalised transverse magneto-optical signal $\Delta I/I$ in saturation as a function of the disc diameter $D$. The wavelengths of incident light is $532$~nm. Experimental data for $15$~nm and $25$~nm thick discs are denoted by squares and dots, respectively. The curves represent theoretical data for the experimental geometries. All data are normalised to their respective values at $D=800$~nm. Theoretical lateral distribution of the in-plane (b.i) and out-of-plane (b.ii) components of polarisation and their relationship (b.iii) as a function of the disc diameter $D$. The polarisation directions are defined with respect to the disc plane. The disc thickness is $5$~nm. Adapted from \cite{Ber15}.} 
\label{fig7}
\end{figure}

It was observed that $\Delta I/I$ is larger for smaller discs, which is shown in Fig.~\ref{fig7}(a) where the Kerr effect at the maximum applied field strength is plotted as a function of the disc diameter $D$. The same trend was confirmed by simulations. This result is counter-intuitive because the magnetic material of the discs is the same and thus all discs should produce a magneto-optical effect of the same strength. At the operating wavelength $532$~nm the values of $\Delta I/I$ are unaffected by plasmon resonances and/or diffraction phenomena. Therefore, the behaviour observed in Fig.~\ref{fig7}(a) may be attributed only to the electric field and polarisation patterns in the discs.

This explanation is supported by simulations [Fig.~\ref{fig7}(b)]. The origin of the dependence of $\Delta I/I$ on $D$ becomes apparent from the relationship between the out-of-plane and in-plane polarisation components (defined with respect to the disc plane). The in-plane component is parallel to the electric field of incident light. It is the largest polarisation component that depends on the optical properties of the disc. In contrast, the out-of-plain polarisation component is induced magneto-optically. 

One observes that in the centre of the discs, the relationship between the out-of-plane and in-plane polarisation components display a response of infinite film, i.e. it does not depend on the disc size. However, a bright ring-like structure at the edges of the disc is substantially larger, and it is responsible for the signal increase in disc with small diameters.    

\section{Applications of magneto-plasmonic nanoantennas}

\subsection{Magneto-plasmonic rulers and biosensors}

In this section, we discuss the application of magneto-plasmonic nanoantennas as the building blocks of novel plasmon rulers and biosensors. The operation of plasmon rulers relies on the near-field-zone coupling between the plasmon modes of two or more adjacent nanoantennas \cite{Liu11}. The strength of the coupling strongly depends on the gap size between the nanoantennas, which leads to high sensitivity of light scattering to the gap size. This effect allows measuring distances at the nano-scale with high accuracy, which is characterised by a figure-of-merit defined as the relationship between the sensitivity of the nanoantenna and the width of the plasmon resonance peak.

Rulers may be useful in material and life sciences. Usually, optical tools used to measure distances at the nano-scale rely on F\"{o}rster resonance energy transfer (FRET) spectroscopy and the use of fluorophores \cite{Shr15}. However, these tools suffer from photobleaching, which must be eliminated because it can alter the measured value of the resonance energy transfer process. This drawback is not present in plasmonic rulers, which allows employing them to complement and improve the existing technologies.

\begin{figure}[t]
\centering\includegraphics[width=8.5cm]{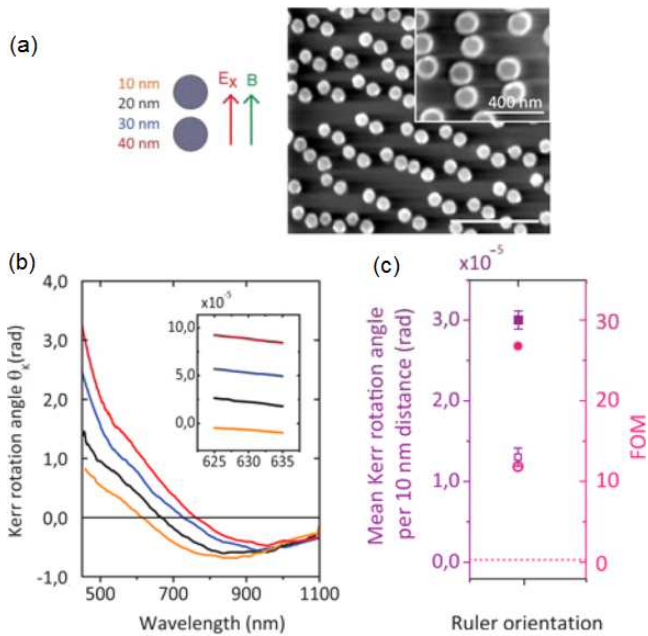}
\caption{ (a) Configuration of the most sensitive ruler showing the orientation of the electric and magnetic fields, and a scanning electron microscopy image of the nickel ruler with the $30$~nm gap size. (b) Spectra of the Kerr polarisation rotation angle for the magneto-plasmonic rulers with different gap sizes (the same colours as in Panel (a) are used). The inset shows the zoomed view. (c, left axis) Mean Kerr rotation angle per $10$~nm for the ruler configuration in Panel (a) and (c, right axis) the corresponding FOM. Solid symbols: $10-30$~nm nano-gap range, open symbols: $10-40$~nm nano-gap range regime. The horizontal dotted line: estimated FOM of the non-magnetic plasmon rulers. Adapted from \cite{Zub15}.} 
\label{fig6x}
\end{figure}

A novel plasmonic ruler consisting of disc magneto-plasmonic nanoantennas [Fig.~\ref{fig6x}(a)] was proposed in \cite{Zub15}. Two batches of nanoantennas were fabricated using nickel and cobalt. The use of these materials in conventional non-magnetic rulers would be challenging because the plasmon resonances of individual all-ferromagnetic nanoantennas are damped by light absorption in the metal. Moreover, when the gap between the adjacent nanoantennas is small, the mode hybridisation leads to the appearance of broad resonance line shapes in the spectra of the ruler.  When nickel is used, the resulting line shapes are just marginally sensitive to the variation in the gap size, which is already unacceptable for the ruler operation. In cobalt rulers, the same resonance features become even less pronounced \cite{Zub15}.

However, the magneto-plasmonic ruler can report nano-gap distances via a different light-matter interaction mechanism -- the Kerr polarisation rotation effect that appears in the presence of technologically attainable external static magnetic fields \textbf{B}. In \cite{Zub15} the spatial orientation of these fields with respect to the ruler was controlled to compare different operation regimes. The Kerr rotation angle spectrum of the most sensitive ruler configuration [Fig.~\ref{fig6x}(a)] is shown in Fig.~\ref{fig6x}(b) for the different gap distances from $10$~nm to $40$~nm. The capability of this ruler to resolve the gap size is better seen in the inset in Fig.~\ref{fig6x}(b), which shows the zoomed $625-635$~nm spectral range available in the most common single-wavelength magneto-optical Kerr effect (MOKE) set-ups.

The advantage of the magneto-plasmonic ruler over its non-magnetic counterpart is summarised in Fig.~\ref{fig6x}(c), the left axis of which demonstrates how the most sensitive ruler configuration is compared against the mean Kerr rotation angle per $10$~nm distance in the nano-gap. One can see that both high absolute rotation per distance and the smallest error in the mean rotation variation are achievable in the $10-30$~nm and $10-40$~nm nano-gap ranges. This implies that the magneto-plasmonic ruler can measure both small and large distances with the same precision, which is difficult to achieve with non-magnetic plasmon rulers. Moreover, the FOM of the magneto-plasmonic ruler [Fig.~\ref{fig6x}(c, right axis)] is $\sim 50$ times larger than that of the non-magnetic plasmon rulers ($\sim 0.62$, marked with the horizontal dotted line).

Next, we discuss the application of magneto-plasmonic nanoantennas in ultra-sensitive devices for label-free moleculat-level detection. In sensors based on non-magnetic nanoantennas, the resonantly enhanced electromagnetic fields near the metal surface allow for probing extremely small changes in the surrounding environment with high sensitivity \cite{Ang10}. Nevertheless, this sensitivity remains insufficient for many real-life applications, which motivates research on alternative designs.

To better understand the alternative design options one should recall the operating principles of biosensors based on surface plasmon resonances (SPR's) \cite{Bar03}. SPR's are charge density oscillations that generate highly localised electromagnetic fields at the interface between a metal and a dielectric. The excitation condition of SPR's strongly depends on the refractive index of the dielectric medium, which is the operating principle of detection of the SPR biosensors.

The sensitivity of SPR sensors improves when a non-magnetic/ferromagnetic metal multilayer is used instead of the single noble metal and a static magnetic field is applied \cite{Sep06}. In this case, SPR's are combined with the magneto-optical activity of the magnetic multilayer, which is the effect responsible for the improved sensitivity.

By analogy with SPR-based sensors, by substituting non-magnetic nanoantennas with magneto-plasmonic ones it should be possible to additionally improve the sensitivity. A realisation of this idea was demonstrated in \cite{Mac15} by using magneto-plasmonic nanoantennas consisting of nickel discs fabricated on top of a glass substrate. In contrast to conventional nanoantenna-based sensors, the nickel nanoantennas were designed such that they produce an exact phase compensation in the electric field component of otherwise elliptically polarised transmitted light at a specific wavelength $\lambda_{\rm{\epsilon}}$. Under this condition, a vanishing ellipticity $\epsilon$ (i.e. the so-called $\epsilon$ null-point corresponding to linear polarisation) is produced at $\lambda_{\rm{\epsilon}}$. Importantly, light polarisation changes can be measured with high precision. Thus, the determination of $\lambda_{\rm{\epsilon}}$ provides a phase-sensitive identification of the plasmon resonance of the magneto-plasmonic nanoantenna. 

\begin{figure}[t]
\centering\includegraphics[width=8.0cm]{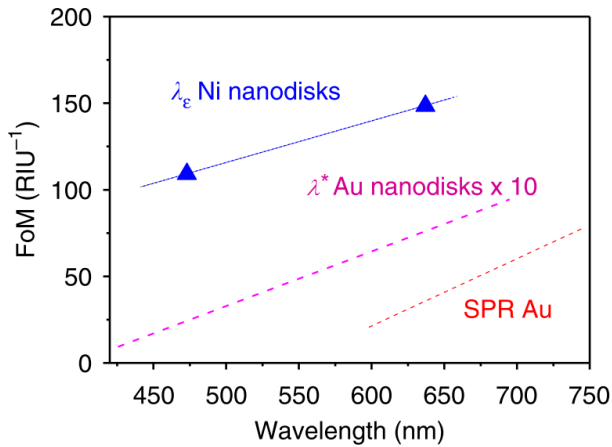}
\caption{ Sensing FOM of nickel nanoantennas $[(\Delta \lambda_{\rm{\epsilon}} / \Delta n)/\rm{FWHM}]$ (markers, the blue line is the guide to the eye) as compared with gold nanoantennas $[(\Delta \lambda^{*} / \Delta n)/\rm{FWHM}]$ and gold SPR-based sensor, in the spectral range $420-750$~nm. FWHM stands for the full width at half maximum, RIU stands for refractive index unit, and SPR means surface plasmon resonance. Adapted from \cite{Mac15}.} 
\label{fig6x2}
\end{figure}

The bulk refractive index sensitivity $S_{\rm{RI}}=\Delta \lambda^{*} / \Delta n$, being $\Delta \lambda^{*}$ the shift in the plasmon resonance position $\lambda^{*}$ in nanometres over the change in the environmental refractive index $\Delta n$,  was measured to quantify the sensing performance of the device immersed into solutions with different indices of refraction $n$. Figure~\ref{fig6x2} shows a comparison of the sensitivities of gold and nickel disc nanoantennas on glass. One can see a several orders of magnitude improvement of the FOM of nickel nanoantennas as compared with gold nanoantennas.

\subsection{Non-reciprocal and one-way devices} \label{Non-reciprocal plasmonic nanoantennas and one-way optical waveguides}

Light propagation is usually reciprocal. However, it is also required to have optical components that exclude undesirable light. Such components are indispensable in optical communication technology where they may serve as optical isolators, gyrators and circulators \cite{2, Sha14}.

Usually, the operation of such devices is based on non-reciprocal optical phenomena that occur in magneto-optical materials. However, the magneto-optical response of naturally occurring materials is weak \cite{2} and may not always be adequate for applications in integrated optical circuits and telecommunications because the applied magnetic fields have to be large.

Surface magneto-plasmons are also known to exhibit non-reciprocal propagation properties \cite{Arm13}. However, the non-reciprocity effects observed in all-ferromagnetic films and multilayers consisting of ferromagnetic and noble metal layers are also weak and require large magnetic fields \cite{Arm13}. 

An enhancement of non-reciprocal effects has been demonstrated in magneto-plasmonic nanostructures such as gratings and resonators \cite{Arm13, Mak15}. Here we are interested in the application of non-reciprocal guided plasmon modes in plasmonic nanoantennas and similar structures that exhibit one-way propagation properties.

Our discussion starts with the extraordinary optical transmission (EOT) through magnetised metallic subwavelength hole arrays. Conventional EOT is the phenomenon of enhanced transmission of light through a subwavelength aperture in an opaque metallic film. The classical aperture theory predicts that incident light of a certain wavelength will be evenly diffracted by a subwavelength aperture in all directions and the far-field transmission will be small. In EOT, however, the regularly repeating nanostructure enables a several orders of magnitude larger transmission efficiency than that predicted by the classical aperture theory \cite{Ebb98}. 

The EOT effect was also investigated in magnetised nanostructures \cite{Str99}. When the wavelength of incident light is larger than the hole size and the array period, it was found that the frequency of the transmission peak depends strongly on both the magnitude and the direction of the applied in-plane magnetic field. This idea was further developed in \cite{Hel05, Bat07, Kha07, Str08, Zho09, Str14}.

\begin{figure}[t]
\centering\includegraphics[width=8.5cm]{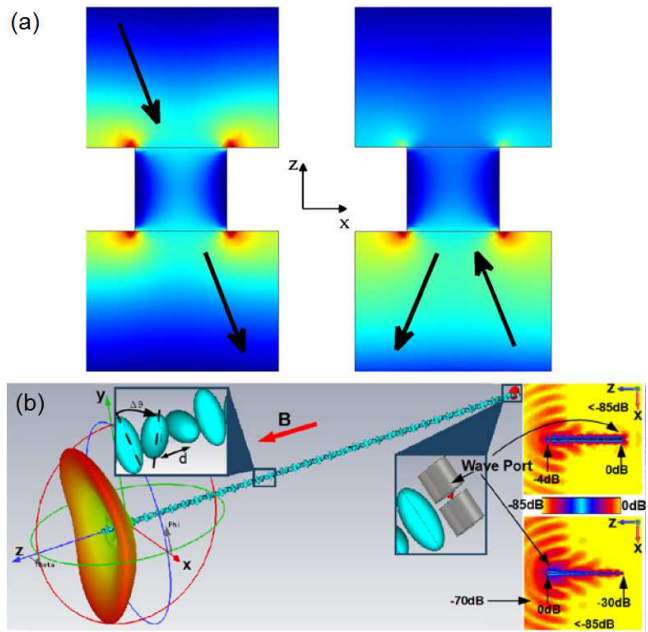}
\caption{ (a) Field profiles corresponding to two opposite directions of incidence (shown by arrows) on a PEC perforated film with an asymmetrically magnetised cladding. Adapted from \cite{Kha10}. (b) Matched magneto-plasmonic nanoantenna. Main panel: General view with a typical far-field pattern. The chain is excited by a quantum emitter that generates a guided mode that propagates along the positive \textit{z}-axis. There is no back reflections at the far end of the nanoantenna because this mode is converted to radiation. Insets: Near-field profiles of the \textit{E}-field. Upper inset: Energy is transmitted from the input port to the far end with minimal losses that occur due to light absorption in the material. Lower inset: Due to the one-way property, if the port is located on the other end the chain, the excitation efficiency is low. Adapted from \cite{Had12}.} 
\label{fig7x1}
\end{figure}

In addition, a new concept of non-reciprocal spoof surface plasmons (NSSPs) in EOT structures was introduced in \cite{Kha10}. It was demonstrated that, by breaking the time-reversal symmetry, NSSPs enable the so-called one-way EOT effect. Figure~\ref{fig7x1}(a) shows the out-of-plane dynamic magnetic field distribution in an optically thick perforated perfect electric conductor (PEC) film sitting on top of an in-plane magnetised insulating substrate. The superstrate is non-magnetic, which implies that the PEC film is surrounded by an asymmetrically magnetised cladding. A strong near-field localisation near the square nano-holes in the PEC film is due to the excitation of leaky NSSPs. One can see that for the forward incidence a nearly complete transparency of the structure is achieved. However, for the backward incidence the transmission through the structure is low. The same effect was also predicted for a structure in which the PEC material was replaced by gold.

One-way waveguiding structures consisting of plasmonic nanoparticles have also attracted considerable attention \cite{Had10, Maz12, Maz14}. We focus on systems that consist of a linear chain of magneto-plasmonic nanoparticles placed into an external magnetic field, which can be applied either in the transverse or longitudinal direction with the respect to the light propagation direction in the nanoparticle chain. In the longitudinal configuration, the chain exhibits optical Faraday rotation. If the magnetised nanoparticles are plasmonic ellipsoids arranged as a spiral, the interplay between the Faraday rotation and the geometrical spiral rotation (structural chirality) strongly enhances non-reciprocity \cite{Had10}. Because of this interplay, in the resulting waveguide light propagates in one direction only. A similar effect was also demonstrated in the transverse magnetic field configuration \cite{Maz12}.

Strongly non-reciprocal and one-way nanostructures can also be used to form metasurfaces. In contrast to one-way nanoparticle chains, in metasurfaces light is manipulated in two dimensions \cite{Maz14}. The resulting surfaces -- the metaweaves -- possess generalised non-reciprocity such as the sector-way propagation. Different metaweaves designs and their properties were discussed in \cite{Maz14}.     

Next we discuss potential applications of the highlighted one-way structures. Impedance matching of nanophotonic circuits with plasmonic nanoantennas is a difficult problem \cite{Alu08}. This problem may be solved by employing one-way waveguides in which back-reflection is not present and the feed signal is converted into the optical antenna radiation with high efficiency. Therefore, it was proposed that a terminated one-way waveguide [Fig.~\ref{fig7x1}(b)] may serve as a complex device consisting of a waveguiding section, an element with matching functionality, and an antenna \cite{Had12}. It was shown that the non-reciprocal operation results in two different emission and direction patterns. In addition, simulations demonstrate a significant dynamic beam scanning functionality of the device.

\begin{figure}[t]
\centering\includegraphics[width=8.5cm]{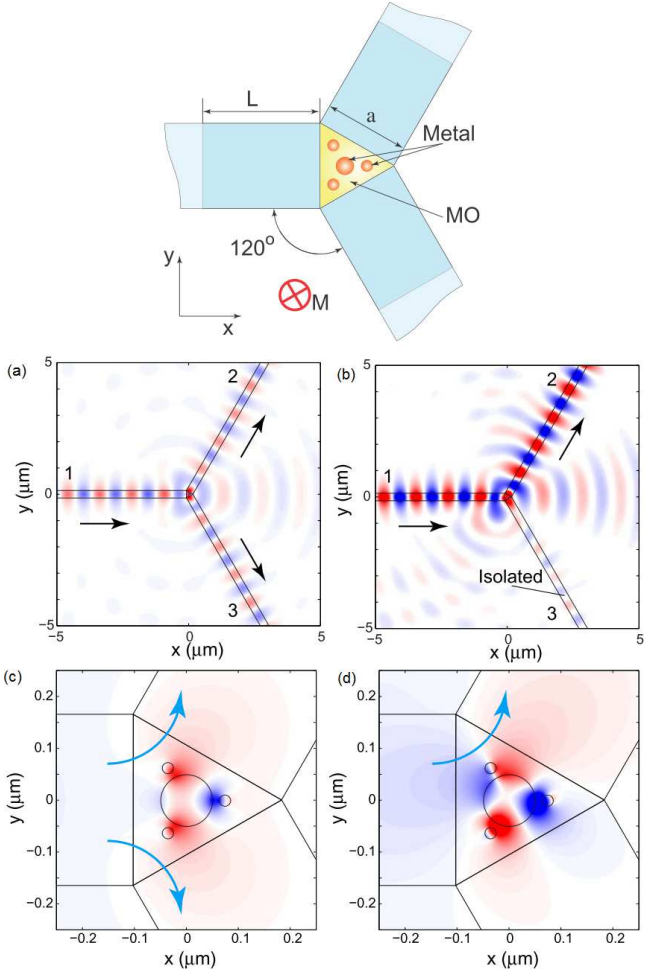}
\caption{ Magneto-plasmonic circulator based on a structure consisting of a non-magnetic metal nanorod with radius $R = 50$~nm surrounded by three other metallic nanorods with radii $R = 10$~nm. The distance between the centres of the thinner and thicker nanorods is $80$~nm. The nanorods are embedded into a magneto-optical (MO) material -- bismuth iron garnet (BIG) magnetised along the \textit{z}-axis. Simulation results for the distribution of the \textit{z}-oriented optical magnetic field $H_{\rm{z}}$ in the circulator. The external static magnetic field is 'off' in Panel (a) and 'on' in Panel (b), respectively. Arrows show the direction of the power flow. Numbers $1-3$ label the ports of the circulator. The free-space wavelength of incident light is $1430$~nm. (c, d) Snapshot of the near-field $H_{\rm{z}}$ profiles corresponding to Panels (a) and (b), respectively. The arrows show schematically the coupling between input port and the two output ports. Adapted from \cite{Dav13_1}.} 
\label{fig7x2}
\end{figure}

So far plasmon-assisted enhancement of magneto-optical activity has been observed in the far-field zone response of nanostructures. However, in \cite{Dav13} it was demonstrated that plasmon excitations also enable non-reciprocal effects in the near-field zone, which may be useful in ultra-small magneto-plasmonic circulators \cite{Dav13_1}. Figure~\ref{fig7x2} shows the results of simulations of the proposed device, which consists of a non-magnetic metal nanorod antenna with radius $R = 50$~nm surrounded by three other metal nanorods with radii $R = 10$~nm. The distance between the centres of thinner and thicker nanorods is $80$~nm. The nanorods are embedded into a magneto-optical material -- bismuth iron garnet (BIG). The static magnetic field is orientated along the \textit{z}-axis and the wavelength of incident light in free space is $1430$~nm. When the magnetic field is turned off [Fig.~\ref{fig7x2}(a)], one observes a symmetric beam splitting into the two arms of the device. However, in the presence of the external magnetic field in Fig.~\ref{fig7x2}(b) one observes a substantial breaking of the symmetry between the power output at ports $2$ and $3$. The power is predominantly guided into the arm $2$, while the other arm carries much less power.

Near-field-zone field distributions [Fig.~\ref{fig7x2}(c, d)] were investigated to understand the mechanism of this strong symmetry breaking. When the external magnetic field is absent, the near-field pattern [Fig.~\ref{fig7x2}(c, d)] is symmetric, which ensures a strong and symmetric coupling between the junction arms and the magneto-plasmonic structure. However, when the external magnetic is present the near-field radiation is tilted and rotated, which leads to the change of coupling between the structure and the junction arms \cite{Dav13_1}.

\section{Photoacoustic imaging with magneto-plasmonic nanoparticles} \label{imaging}

Scientific breakthroughs from physics, chemistry, biology and medicine have led to the development of biomedical imaging techniques with high sensitivity and resolution. These technique are essential for both understanding of biological phenomena and detection of diseases. The representative imaging modalities include magnetic resonance imaging (MRI), computed tomography (CT), positron emission tomography (PET), optical fluorescence imaging, ultrasound (US) imaging, and photoacoustic (PA) imaging \cite{Mae09, Wan10, Wei14, Shi15}. Some of these techniques rely on the interaction of light with biological tissues and artificial contrast agents such as magneto-plasmonic nanoparticles, which warrants their discussion in this separate section.

\begin{figure*}[t]
\centering\includegraphics[width=15cm]{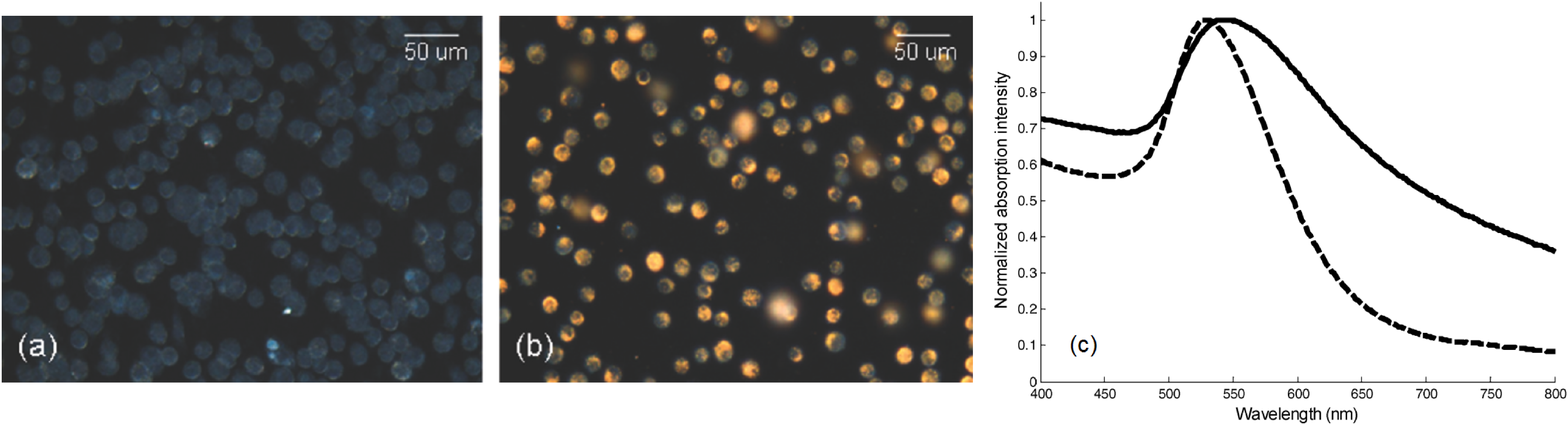}
\caption{ Dark-field reflectance optical images of intact macrophage cells (a) and cells loaded with gold nanoparticles (b). (c) Extinction spectra of cells loaded with gold nanoparticles (solid line) and gold nanoparticles only (dashed line). Both absorption spectra are normalised to their corresponding maxima. Adapted from \cite{Wan09}.} 
\label{fig9}
\end{figure*}

Ultrasound is a valuable imaging modality and diagnostic tool used in clinical practice. Ultrasound has several advantages over the other imaging techniques such as real-time imaging, high resolution, good penetration depth, cost effectiveness, and portability. These advantages have led to the development of different ultrasound imaging modalities, including intravascular ultrasound (IVUS) imaging \cite{Wan10}.

In PA imaging, a laser beam is used to excite ultrasonic waves, which, in general, can be generated through different physical mechanisms that include thermal expansion, vaporisation, photochemical  processes, and optical breakdown \cite{Wan10}. However, in biomedical applications of PA imaging, the only biologically safe mechanism is thermal expansion. In this case, biological tissues absorb laser light and generate a broadband ultrasound signal, which is detected using a transducer that converts acoustic waves into electric signals. These signals are processed to produce a PA image.

Unfortunately, thermal expansion is one of the least efficient mechanisms of light-ultrasound interaction. Consequently, it produces ultrasound of relatively low amplitude. On the other hand, contrast in PA imaging arises from the natural variation in the optical absorption of tissue components. The absorption cross-section of plasmonic nanoparticles is many orders of magnitude higher than that of tissues. Consequently, recent attention has turned to the use of plasmonic nanoparticles as contrast agents for PA imaging \cite{Wan09, Hom12, Shi15}. Figure~\ref{fig9} shows the results of measurements of the plasmonic nanoparticle impact on the identification of macrophage cells, which play an important role in atherosclerosis \cite{Moo13}. Cells were loaded with $50$~nm-diameter spherical gold nanoparticles [Fig.~\ref{fig9}(a, b)]. Optical absorption spectra of cells loaded with nanoparticles and nanoparticles along are presented in Fig.~\ref{fig9}(c). Compared to the spectrum of gold nanoparticles (dashed curve), the spectrum of cells loaded with the nanoparticles (solid curve) is red-shifted and its line-shape is also broadened. The shift and broadening can be explained by the cumulative effect of plasmon resonance coupling of adjacent gold nanoparticles after being internalised by macrophages.

However, whereas the combined dual-modality US/PA imaging benefits from plasmon-assisted high contrast of its PA component and ultrasound provides morphological details of the anatomy, ambiguities arising from PA background signals remain in resulting images \cite{Shi15}. A strong background signal is a common challenge regardless of the imaging modality. For example, in optical imaging a background signal originates from scattering, absorption and autofluorescence that obscure specific signals from targeted contrast agents. Similarly, in magnetic resonance imaging the background signal originates due to high water content of the human body.

\begin{figure}[t]
\centering\includegraphics[width=8.0cm]{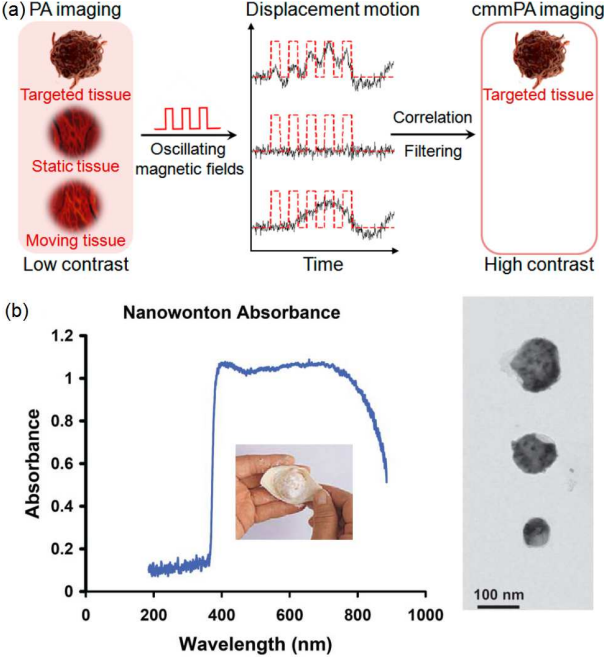}
\caption{ (a) Schematic of tumour detection by means of cyclic mmPA imaging. Contrast enhancement of tumour/background is achieved by suppressing both signals from static tissues and regions with random motions from moving tissues not related to the frequency of oscillating magnetic fields. Specific tumour signals from targeted tissues are coherently responsive to cyclic magnetic motions. Adapted from \cite{Li15}. (b) Absorption spectrum of a nanowonton with the medium peak wavelength around $700$~nm (left) and transmission electron microscopy image of three nanowontons in various diameter (right). The inset in the right panel shows a real wonton. Adapted from \cite{Bou08}.} 
\label{fig10}
\end{figure}

To address the problem of background noise in PA imaging, a new imaging modality called magneto-photo-acoustic imaging (also called magneto-motive photoacoustic (mmPA) imaging) has been introduced \cite{Gal09, Jin10, Qu11, Li15}. As the name suggests, this technique exploits magnetic nanoparticles as the PA contrast agent. In addition, an external pulsed magnetic field is applied during PA image acquisition. Because of this field magnetic nanoparticles create a vibrating motion. However, non-magnetic biological tissues and liquids remain unaffected by the magnetic field. By detecting the motion of the nanoparticles, PA signals from the nanoparticles can be distinguished from those originated from the background [Fig.~\ref{fig10}(a)]. This functionality has been demonstrated using Fe$_3$O$_4$ nanoparticles and magnetic-gold core-shell nanoparticles \cite{Jin10, Li15}, Fe$_3$O$_4$-gold nanorods \cite{Qu11}, and gold-coated cobalt nanoparticles \cite{Bou08}. 

Previously, oxidation-induced instability and toxicity of cobalt nanoparticles have prohibited their wide application as a contrast agent. However, the gold coating proposed in \cite{Bou08} solves this problem. By analogy with Chinese food called wonton, these cobalt-gold nanostructures are called nanowontons [Fig.~\ref{fig10}(b)]. They have a cobalt core and a gold thin-film coating, and are constructed similar to edible wontons. The thickness and the shape of the gold layer of nanowontons can be engineered to control the absorption spectral range at will, which may be useful to match the near infrared laser excitation wavelength $700$~nm used in PA imaging for the optimisation of the photothermal response [Fig.~\ref{fig10}(b)].

\section{Future directions and outlook}

Clearly, reception, emission and, more generally, control of light at the nanoscale are central to a number of emergent nano-optical applications. Nanoscale magneto-plasmonics is one of the technologies that has the potential to offer such functionality. Although magneto-plasmonic nanoantennas can operate similar to conventional non-magnetic plasmonic nanoantennas having a wide range of unique applications \cite{Bra09, Alu09, Nov11, Gia11, Bon12, Ber12, Agi12, Bia12, Mak12, Che12, Kra13, Alu13, Agi13, Cha13, Mon14, Bor14}, they may offer more flexibility and provide better performance than their non-magnetic counterparts. This is because the operation of magneto-plasmonic nanoantennas additionally relies on magneto-optical and magnetic properties of nanostructured ferromagnetic metal and magneto-insulating materials. Rich physics and mature technology of magnetic nanostructures (recall computer hard drives, non-volatile magnetic random access memory, magneto-optical discs, etc. \cite{2, Hir14}) generously contribute to the functionality of magneto-plasmonic nanoantennas and thus considerably increase their application options. Furthermore, some experimental techniques traditionally used to characterise magnetic nanomaterials now find applications in studies of purely optical properties of nanostructures \cite{Sta15}. 

We would like to conclude this review by indicating an emergent research direction in the area of magneto-plasmonic nanoantennas -- nonlinear magneto-plasmonic nanoantennas.

Nonlinear optical effect have been used in spectrally tunable plasmonic nanoantennas \cite{Agi13, Kra13}. In addition, it has been demonstrated that the second harmonic generation (SHG), third harmonic generation (SHG), and four-wave mixing (FWM) nonlinear optical effects can be enhanced by exploiting a strong local field localisation in nanonatennas. This enhancement has opened novel opportunities for the development of integrated on-chip photonic devices for applications in telecommunications, signal processing, precision measurements, and imaging. 

In principle, magneto-plasmonic nanoantennas may also be used to achieve spectral tuning or enhance nonlinear effects. However, their advantage over non-magnetic nanoantennas is that they can additionally exhibit nonlinear magneto-optical properties. For example, in \cite{Kru13, Raz13, Zhe14, Raz15} it was shown that the nonlinear magneto-optical Kerr effect and the SHG effect can be enhanced at the frequencies of plasmon modes supported by arrays of ferromagnetic metal nanorods or magnetic multilayer heterostructures.

Last but not least, the inverse magneto-optical Faraday \cite{Pop94, Kim05, Per06} and transverse Kerr \cite{Bel12} effects are also nonlinear, which was demonstrated in both continuous ferromagnetic metal films and nanostructures. The inverse magneto-optical effects are important for ultra-fast all-optical control of magnetisation, which may be useful in data storage systems and generation of spin waves with light pulses \cite{Per06}.    

\section{Acknowledgements}

This work was supported by the Australian Research Council (ARC) through its Centre of Excellence for Nanoscale BioPhotonics (CE140100003). The author would like to thank Prof. M. Kostylev and Dr. P. Metaxas for valuable discussions preceding the creation of this work.


\end{document}